# Stress-Induced Intercalation Instability

Youtian Zhang[1] and Ming Tang[1*]

1. Department of Materials Science & NanoEngineering, Rice University, Houston, TX 77005, USA.

*Corresponding author. Email address: mingtang@rice.edu

**Abstract**

We present a linear stability analysis to demonstrate that a flat coherent phase boundary formed by the (de)intercalation of solutes into a compound is unstable against perturbations with wavelengths larger than a critical wavelength. This critical wavelength is controlled by the competition between the interface energy and the elastic strain energy caused by the misfit between the solute-rich and solute-poor phases. It increases with the distance between the phase boundary and free surface of the compound, and so the instability is most pronounced when the boundary is close to the surface at the early stage of the (de)intercalation process. Numerical calculations show that such instability leads to non-uniform intercalation behavior. We find that uniform intercalation cannot be achieved unless the phase boundary moves at a speed greater than a critical velocity. Estimate of the magnitude of this velocity suggests that the stress-induced intercalation instability is generally operative in intercalation compounds used for battery applications.

Keywords: interface, morphological stability, internal stress, intercalation compounds, Li-ion battery

# 1. Introduction

Stress plays an important role in microstructure evolution in solid materials. A well-known example is the stress-driven morphological instability during epitaxial thin film growth[1-3], where the coherency stress generated by the film/substrate misfit destabilizes the planar film surface and triggers a film-to-island transition in the growth behavior, see Figure 1a. Now often called the Asaro-Tiller-Grinfeld (ATG) instability, it was first analyzed by Asaro and Tiller[4], Grinfeld[5] and Srolovtiz[6] for a semi-infinite solid and later extended by Spencer et al.[7] and Freund and Jonsdottir[8] to an epitaxially strained film on substrate. Similar instabilities resulting from the competition between stress and interface tension were also observed in other types of systems [9, 10].

This work concerns the stability of interfaces in intercalation compounds, which are materials with a host matrix that can reversibly accommodate foreign ions, atoms or molecules. In recent years, this group of materials has received tremendous interest for energy storage applications including lithium-ion battery electrodes (e.g. graphite and layered transition metal oxides) and hydrogen storage medium (e.g. metal hydrides). Many intercalation compounds undergo one or more first-order phase transformations upon solute insertion or extraction often accompanied by appreciable volume change. Large coherency stress usually arises during the intercalation process. Here we predict a phenomenon analogous to the ATG instability in intercalation materials, where a flat phase boundary between the parent and product phases is destabilized by the misfit stress, resulting in non-uniform (de)intercalation behavior, see Figure 1b. Such instability may be considered as an "inverted" ATG instability as it is the buried interface between the product ("film) and parent ("substrate") phase rather than the surface that becomes unstable with respect to perturbations. We suggest that such instability provides a plausible explanation to recent observations of jagged (de)intercalation fronts (see Figure 1c and 1d) in lithium-ion electrodes[11-13] and is potentially relevant to a large number of intercalation compounds with broad implications for their performance and degradation in battery applications.

In this paper, we use linear stability analysis and phase-field simulations to reveal the essential features of the stress-induced intercalation instability and the subsequent phase evolution. Criteria for maintaining uniform intercalation throughout the intercalation process are derived. Despite its similarity to the ATG instability, the stress-induced intercalation instability possesses a few notably distinct features. While the ATG instability is enhanced by increasing epitaxial film thickness on the substrate, the interface instability within a stressed intercalation body exhibits the opposite trend and is continuously weakened by the growth of the product phase. It is possible that a planar phase boundary becomes stable against any perturbations when it is sufficiently far away from the free surface of the system. Another interesting finding is that large intercalation flux and sluggish diffusion of the intercalating solutes help stabilize the interface, which is contrary to the role diffusion often plays in promoting interface instability, e.g. dendrite growth during solidification[14, 15].

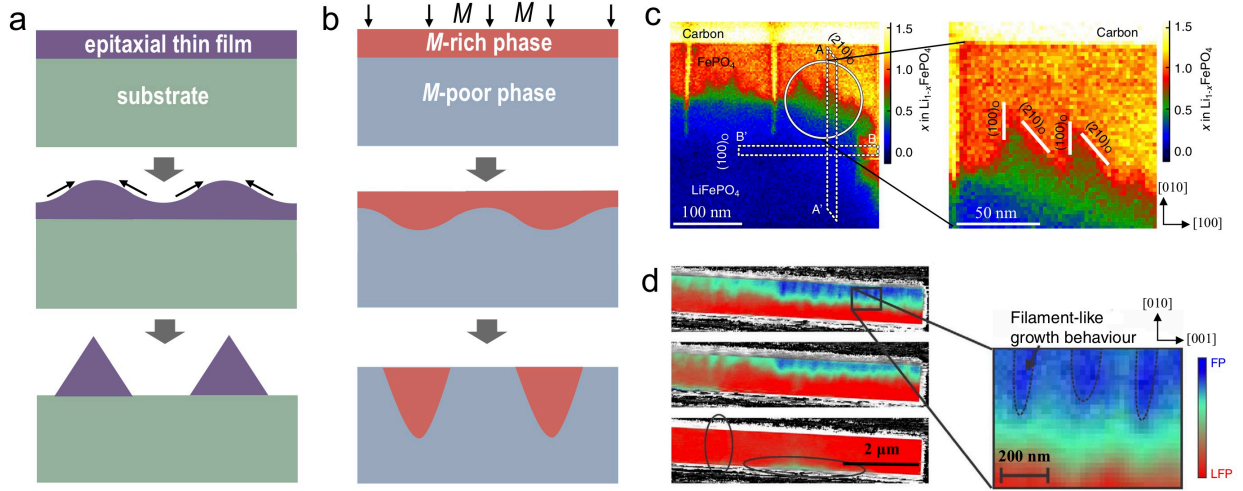

**Figure 1.** a) Schematic of the ATG instability during epitaxial thin film growth that leads to island growth. b) Schematic of the stress-induced intercalation instability predicted by this work, where the planar (de)intercalation front is destabilized by the coherency stress due to the misfit between the solute-rich and solute-poor phases. c) Presence of jagged $LiFePO_4$ / $FePO_4$ phase boundaries in a chemically delithiated $LiFePO_4$ particle. Adapted from Ref. [12] with permission. d) Observation of wavy phase boundaries in an electrochemically cycled $LiFePO_4$ single crystal. Adapted from Ref. [11] with permission.

## 2. Results

### 2.1 Linear stability analysis

We first employ the linear stability analysis to study the stability of a flat (de)intercalation front in a compound, which a solute denoted as *M* can be inserted into or extracted from. Here we consider the intercalation of *M* into a two-dimensional (2D) semi-infinite system under plane strain condition. The analysis results also apply to the deintercalation process. As illustrated in Figure 2, intercalation induces a phase transition from a *M*-poor phase (II) to *M*-rich phase (I), which forms a uniform layer along the surface at $z = 0$ and a coherent phase boundary at $z = z_0$. For simplicity, assume that the compound is linearly elastic with isotropic and homogeneous elasticity and the transformation strain tensor $\epsilon_{ij}^0$ of phase II→I is isotropic in the *x-z* plane, i.e. $\epsilon_{xx}^0 = \epsilon_{zz}^0 = \epsilon_0$ and other elements are zero. We also assume that *M* concentration only has a narrow variation in both phases and so the composition dependence of $\epsilon_{ij}^0$ may be neglected.

We use the Airy stress function to calculate the stress field after a small sinusoidal perturbation $\delta \cos(kx)$ is applied to the phase boundary as shown in Figure 2. Stress components in 2D systems can be expressed in terms of Airy stress function $\Phi(x,z)$: $\sigma_{xx} = \partial_z^2 \Phi$, $\sigma_{zz} = \partial_x^2 \Phi$ and $\sigma_{xz} = \sigma_{zx} = -\partial_x \partial_z \Phi$, with $\Phi$ satisfying the biharmonic equation[16]:

$$(\partial_x^2 + \partial_z^2)^2 \Phi = 0 \qquad \qquad 1)$$

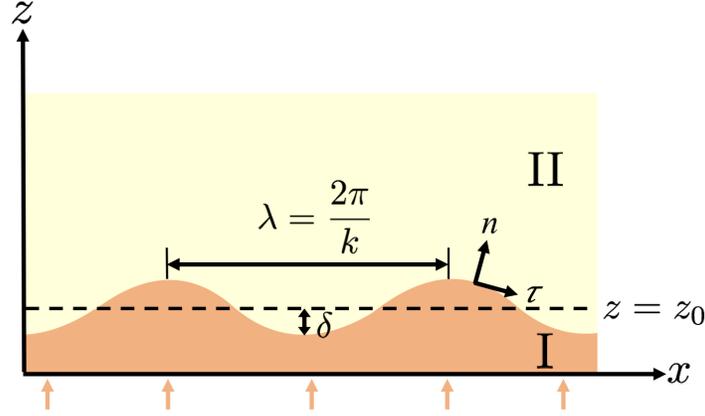

**Figure 2.** Configuration of the intercalation system considered in the linear stability analysis.

When the phase boundary is flat, phase I is uniformly stressed along $x$ with $\sigma_{xx}^I \equiv \sigma_0 = -E\epsilon_0/(1-v^2)$ and $\sigma_{zz}^I = \sigma_{xz}^I = 0$, and phase II is stress-free ($\sigma_{ij}^{II} = 0$). In the presence of an interface perturbation with a wave vector $k$, a general solution to Eq. 1 can be written as:

$$\Phi^I = \frac{\sigma_0 z^2}{2} + (a_1 + b_1 z)\cos(kx)e^{-kz} + (c_1 + d_1 z)\cos(kx)e^{kz} \qquad \text{2a)}$$

$$\Phi^{II} = (a_2 + b_2 z)\cos(kx)e^{-kz} \qquad \text{2b)}$$

where the superscript denotes the phase and $a_{1/2}$, $b_{1/2}$, $c_1$ and $d_1$ are constants to be determined by the boundary conditions of the system. The traction-free boundary condition at the free surface $z = 0$ requires:

$$\sigma_{zz}^I = \sigma_{xz}^I = 0 \qquad \text{3)}$$

At the phase boundary $z = h(x) = z_0 + \delta\cos(kx)$, stress and displacement continuity requires:

$$\sigma_{nn}^I = \sigma_{nn}^{II}, \quad \sigma_{\tau n}^I = \sigma_{\tau n}^{II} \qquad \text{4)}$$

$$u_\tau^I = u_\tau^{II}, \quad u_n^I = u_n^{II} \qquad \text{5)}$$

in which the stress and displacement tensors are expressed in a orthogonal curvilinear coordinate system with $\tau$ and $n$ being the tangential and normal directions of the boundary as shown in Figure 2. In Supplementary Information (SI), we show that Eq. 5 can be replaced by the following boundary conditions:

$$\sigma_{\tau\tau}^I - \sigma_0 = \sigma_{\tau\tau}^{II} \qquad \text{6)}$$

$$(1-v)\frac{\partial \sigma_{\tau\tau}^I}{\partial n} - v\frac{\partial \sigma_{nn}^I}{\partial n} - \kappa\left[(1-v)\sigma_{nn}^I - v\sigma_{\tau\tau}^I + \frac{E\epsilon_0}{1+v}\right] = (1-v)\frac{\partial \sigma_{\tau\tau}^{II}}{\partial n} - v\frac{\partial \sigma_{nn}^{II}}{\partial n} - \kappa\left[(1-v)\sigma_{nn}^{II} - v\sigma_{\tau\tau}^{II}\right]$$

7)

where $\kappa = \delta k^2 \cos(kx)$ is the interface curvature up to the first order of $\delta$, and $E$ and $\nu$ are the Young's modulus and Poisson ratio, respectively. Applying Eqs. 2 to Eqs. 3, 4, 6 and 7 and using the tensor transformation relation

$$\begin{bmatrix} \sigma_{\tau\tau} & \sigma_{\tau n} \\ \sigma_{n\tau} & \sigma_{nn} \end{bmatrix} = \begin{bmatrix} \cos\theta & \sin\theta \\ -\sin\theta & \cos\theta \end{bmatrix} \begin{bmatrix} \sigma_{xx} & \sigma_{xz} \\ \sigma_{zx} & \sigma_{zz} \end{bmatrix} \begin{bmatrix} \cos\theta & -\sin\theta \\ \sin\theta & \cos\theta \end{bmatrix} \qquad 8)$$

we obtain a system of linear equations for the unknown constants $a_{1/2}$, $b_{1/2}$, $c_1$ and $d_1$, from which their values can be calculated. This leads to the following order-$\delta$ expressions of the stress components at the phase boundary:

$$\sigma_{xx}^I - \sigma_0 = \sigma_{xx}^{II} \approx \tilde{\sigma}_{xx}^{int} \equiv \frac{E\epsilon_0}{2(1-\nu^2)} k(3 - 2kz_0 + e^{2kz_0})e^{-2kz_0}\delta \cos(kx) \qquad 9a)$$

$$\sigma_{zz}^I = \sigma_{zz}^{II} \approx \tilde{\sigma}_{zz}^{int} \equiv \frac{E\epsilon_0}{2(1-\nu^2)} k(1 + 2kz_0 - e^{2kz_0})e^{-2kz_0}\delta \cos(kx) \qquad 9b)$$

$$\sigma_{xz}^I = \sigma_{xz}^{II} \approx \tilde{\sigma}_{xz}^{int} \equiv -\frac{E\epsilon_0}{2(1-\nu^2)} k(1 - 2kz_0 + 2e^{2kz_0})e^{-2kz_0}\delta \sin(kx) \qquad 9c)$$

where $\tilde{\sigma}_{ij}^{int}$ is the first-order perturbation to the stress field at the interface, which is continuous across the phase boundary. Eq. 9 shows that the magnitude of the major stress component $\sigma_{xx}$ in phase I is reduced at the "peak" ($h = z_0 + \delta$) and enhanced at the "valley" ($h = z_0 - \delta$) of the perturbed phase boundary. Because stress contributes to the driving force of phase transformation, the less strained peak section of the boundary will move faster than the valley section, therefore destabilizing the initially flat interface. For a quantitative analysis of the perturbation growth rate, we consider two different scenarios below.

*I. Interface-controlled phase transformation*

When the phase II→I transformation is kinetically controlled by interface reaction, the phase boundary velocity can be written as:

$$\frac{dh}{dt} = \frac{dz_0}{dt} + \frac{d\delta}{dt}\cos(kx) = M_I \Delta f \qquad 10)$$

where $M_I$ is the interface mobility and the transformation driving force $\Delta f$ is given by

$$\Delta f = \Delta f_0 - \gamma\kappa + (\tilde{\sigma}_{xx}^{int} + \tilde{\sigma}_{zz}^{int})\epsilon_0 \qquad 11)$$

$\Delta f_0$ is the driving force in the presence of a flat phase boundary, which in the case of battery electrode compounds can be controlled by the external potential applied. Interface perturbation introduces additional contributions $\gamma\kappa$ and $(\tilde{\sigma}_{xx}^{int}+\tilde{\sigma}_{zz}^{int})\epsilon_0$ by interface energy $\gamma$ and stress, respectively. Substituting Eqs. 9 and 11 into Eq. 10 and matching terms in powers of $\delta$, we find:

$$\frac{d\delta}{dt} = M_I\left(f^*ke^{-2kz_0} - \gamma k^2\right)\delta \qquad 12)$$

in which $f^*$ is a characteristic elastic energy density:

$$f^* = \frac{2E\epsilon_0^2}{(1-v^2)} \qquad 13)$$

Eq. 12 shows that $\delta$ grows or decays exponentially with time as $\delta(t)=\delta_0\exp(\omega_I t)$ with the perturbation growth exponent

$$\omega_I = M_I\left(f^*ke^{-2kz_0} - \gamma k^2\right) \qquad 14)$$

Eq. 14 illustrates that coherency stress promotes perturbation growth but interface energy stabilizes a planar boundary. As shown by the dispersion relation $\omega_I(k)$ plotted in Figure 3a, the phase boundary is unstable against any perturbations with wave vectors smaller than a critical wave vector $k_c$, which is given by:

$$k_c = \frac{W(2f^*z_0/\gamma)}{2z_0} \qquad 15)$$

in which $W(\zeta)$ is the product logarithm (or Lambert $W$) function, i.e. $W$ is the principal solution to $\zeta=We^W$. Similar form of $k_c$ also appears in the morphological instability of grain boundary in two-phase coherent solids because of the analogous role of grain boundary in relieving coherency stress [9]. $k_c$ is approximated by $f^*/\gamma(1-2f^*z_0/\gamma)$ at small $z_0$ and $[\ln(2f^*z_0/\gamma)-\ln(\ln(2f^*z_0/\gamma))]/(2z_0)$ for large $z_0$. Although an analytical expression of the wave vector of the fastest growing perturbation $k_m$ is not generally available, it is easy to see that $k_m = k_c/2 = f^*/2\gamma$ at $z_0 = 0$.

Figure 3b shows an important feature of $k_c$, namely, it monotonically decreases with $z_0$ and approaches 0 as $z_0 \to \infty$. This means that perturbations are more and more difficult to amplify when the interface moves away from the surface. To understand this result, one should recognize that an interface perturbation grows only when it reduces the total elastic energy in the system. As such, the presence of a free surface near the phase boundary is important as it relaxes the deformation along the surface normal that is caused by the perturbation. When the phase boundary is far away from the surface, they do not elastically interact with each other and so the free surface effect on stress relief is diminished. In fact, varying the morphology of a phase boundary in an infinite, isotropic elastic body does not effect any change in the elastic energy according to the Bitter-Crum theorem [17], i.e. the boundary is marginally stable against any perturbations. Therefore, a flat phase boundary is most susceptible to instability at the early stage of the intercalation process, when it is close to the free surface. $k_c$ reaches its maximum $f^*/\gamma$ at $z_0 = 0$, and $\lambda_{min} = 2\pi\gamma/f^*$ defines the shortest wavelength of perturbations that may grow during intercalation. To estimate the magnitude of $\lambda_{min}$, we use typical values of $E$ (100 GPa), $\nu$ (0.3), $\epsilon_0$ (0.01) and $\gamma$ (0.1 J/m²) for intercalation compounds and find $\lambda_{min} = 29$ nm. It shows that the unstable perturbations can be accommodated in most battery electrode particles, whose particles sizes are usually larger than 100 nm.

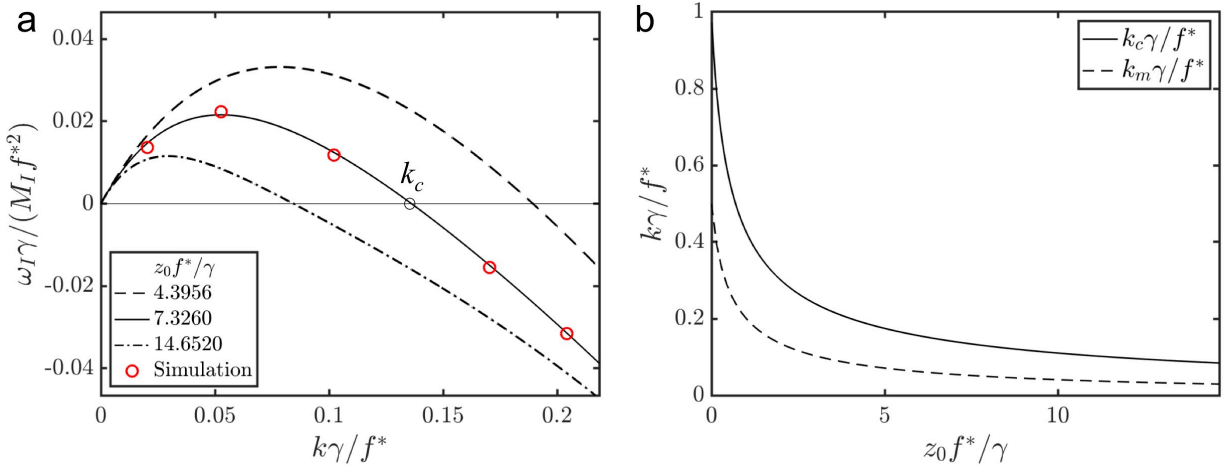

Figure 3. a) Dispersion relation of the perturbation growth exponent $\omega_I(k)$ at different $z_0$ for interface-controlled intercalation. Red circles are $\omega_I$ calculated by phase-field simulations. b) The critical wave vector ($k_c$) and the fastest growing wave vector ($k_m$) as a function of $z_0$. In the plots, $k$ is scaled by $f^*/\gamma$, $\omega_I$ by $M_I f^{*2}/\gamma$, and $z_0$ by $\gamma/f^*$.

## II. Diffusion-controlled phase transformation

Next, we consider the situation where the speed of phase boundary migration is controlled by the diffusion of intercalating solutes in the system. As we assume that both phase I and II only have a small degree of nonstoichiometry, the perturbation-induced spatial variation of $M$ concentration within each phase has an order-$\delta^2$ effect on the stress field and may be excluded from the linear stability analysis. The perturbed stress filed is still described by Eq. 9. Similar to the analysis of the classic Mullins-Sekerka instability [14], we assume that $M$ diffusion in the system is steady-state and satisfies:

$$\nabla^2 \mu^I = \nabla^2 \mu^{II} = 0 \qquad 16)$$

where $\mu^I$ and $\mu^{II}$ are the chemical potentials of $M$ in phase I and II, respectively. In the presence of a flat phase boundary, $\mu^I$ varies linearly with $z_0$ and $\mu^{II}$ is constant. When the interface is perturbed by a sinusoidal wave of amplitude $\delta$, an order-$\delta$ general solution to Eq. 16 is given by:

$$\mu^I = A_1 + B_1 z + (C_1 e^{-kz} + D_1 e^{kz})\delta \cos(kx) \quad \text{17a)}$$

$$\mu^{II} = C_2 e^{-kz}\delta \cos(kx) \quad \text{17b)}$$

where $A_1$, $B_1$, $C_1$, $D_1$ and $C_2$ are constants to be determined from the boundary conditions. Let's consider a constant-flux intercalation process:

$$-M_D \partial_z \mu^I \big|_{z=0} = J \quad \text{18)}$$

where $M_D$ is the mobility of solute $M$. In addition, local equilibrium between phase I and II are satisfied at the phase boundary $z = h(x) = z_0 + \delta \cos(kx)$:

$$\mu^I = \mu^{II} \approx \mu_0 + \gamma\kappa - (\tilde{\sigma}_{xx}^{\text{int}} + \tilde{\sigma}_{zz}^{\text{int}})\epsilon_0 \quad \text{19)}$$

where $\mu_0$ is the equilibrium chemical potential at a flat phase boundary. Note that the sign of the stress term in Eq. 19 is positive for deintercalation. Substituting Eq. 17 into Eqs. 18 and 19 and matching terms in powers of $\delta$, one can solve for the unknown constants $A_1$, $B_1$, $C_1$, $D_1$ and $C_2$. To the first order of $\delta$,

$$\mu^I \approx \mu_0 + \frac{J}{M_D}(z_0 - z) + \left[\left(\gamma\kappa k + \frac{J}{M_D}\right) - f^* k e^{-2kz_0}\right]\frac{\cosh(kz)}{\cosh(kz_0)}\delta \cos(kx) \quad \text{20a)}$$

$$\mu^{II} \approx \mu_0 + (\gamma\kappa - f^* e^{-2kz_0})e^{-k(z-z_0)}k\delta \cos(kx) \quad \text{20b)}$$

The interface velocity can be calculated from the Stefan condition

$$\frac{dh}{dt} = \frac{dz_0}{dt} + \frac{d\delta}{dt}\cos(kx) = -\frac{M_D}{\Delta\rho}\partial_z(\mu^I - \mu^{II})\big|_{z=h(x)} \quad \text{21)}$$

where $\Delta\rho$ is the $M$-molar density difference between phase I and II. Eq. 21 shows that $\delta$ grows or decays exponentially with time, and the perturbation growth exponent is given by

$$\omega_D = \frac{M_D}{\Delta\rho}k\left\{f^* k[1-\tanh(kz_0)] - \gamma k^2[1+\tanh(kz_0)] - \frac{J}{M_D}\tanh(kz_0)\right\} \quad \text{22)}$$

Note that the sign of the last term in Eq. 22 is positive for deintercalation ($J < 0$). Like interface-controlled intercalation, interface perturbations here are also encouraged by stress ($f^*$) and suppressed by interface energy ($\gamma$). However, a new feature is that the third term in Eq. 22, which is proportional to $J/M_D$, also stabilizes the interface. This means that the interface is more stable against perturbations when intercalation flux is large or diffusion is slow. Such behavior draws a sharp contrast to the Mullins-Sekerka instability[14], in which a large supersaturation and small diffusivity in the matrix destabilize the morphology of growing particles during coarsening or solidification. The origin of such difference lies in that the intercalation flux of $M$ arrives at the moving phase boundary from the side of the product phase

(phase I), versus from the side of the parent phase in the Mullins-Sekerka instability. When the boundary is perturbed, its peak segment ($h = z_0 + \delta$) moves away from the free surface, which reduces the diffusion flux arriving at this location and slows down its advancement into the bulk. On the other hand, the valley segment of the perturbed boundary ($h = z_0 - \delta$) receives an increased flux because it is closer to the surface, which causes it to move faster and thus suppresses the growth of the perturbation.

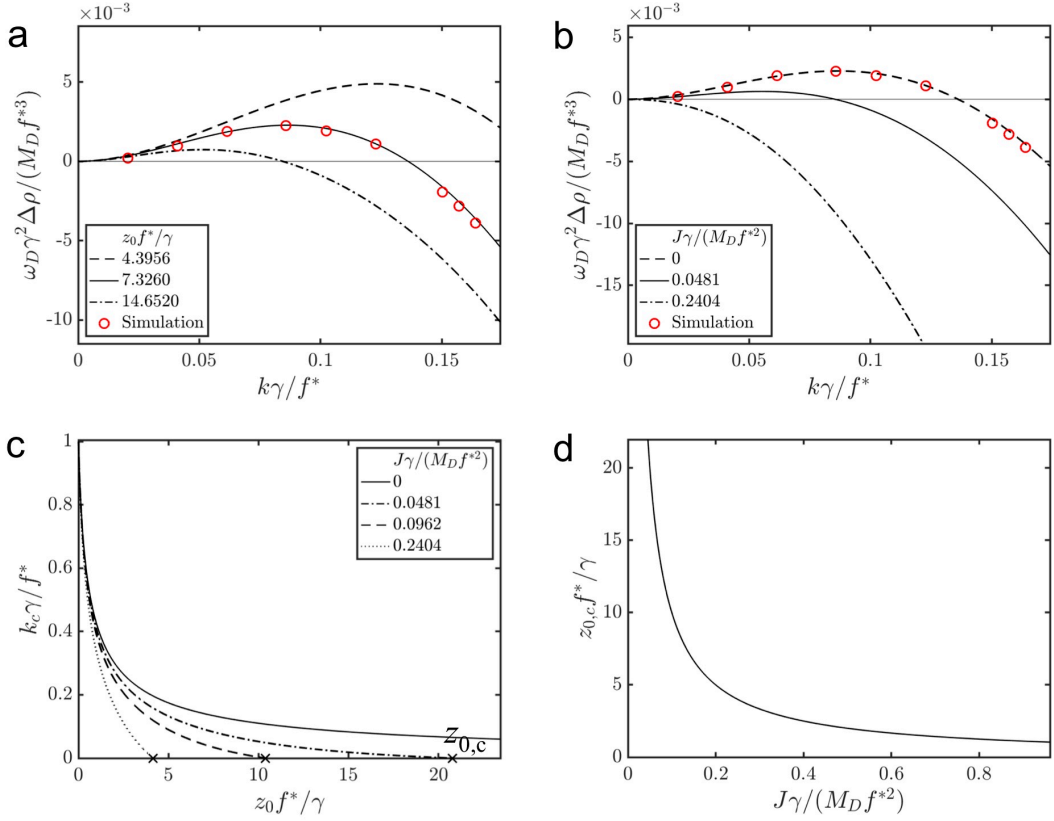

**Figure 4.** Dispersion relation of the perturbation growth exponent $\omega_D(k)$ for diffusion-controlled intercalation when a) $J = 0$ and b) $z_0 f^*/\gamma = 7.326$. Red circles are $\omega_D$ calculated by phase-field simulations. c) Critical wave vector $k_c$ as a function of $z_0$ for different $J$. d) Effect of intercalation flux $J$ on the critical interface location $z_{0,c}$ at which $k_c = 0$. In the plots, $k$ is scaled by $f^*/\gamma$, $\omega_D$ by $M_D f^{*3}/\gamma^2 \Delta\rho$, $z_0$ by $\gamma/f^*$, and $J$ by $M_D f^{*2}/\gamma$.

As shown by the dispersion $\omega_D(k)$ in Figure 4a, the interface is unstable to perturbations with $k$ below a critical wave vector $k_c$. When $J = 0$, i.e. the interface is stationary, $k_c$ is the same as the critical wave vector for interface-controlled intercalation as given by Eq. 15. However, Figure 4b shows that $k_c$ and $\omega_D$ decrease monotonically with $J$. For a non-zero $J$, Figure 4c shows that $k_c$ diminishes to zero when $z_0$ reaches a critical value $z_{0,c}$. The phase boundary is thus absolute stable against perturbations at $z_0 > z_{0,c}(J)$. We can determine $z_{0,c}(J)$ by inspecting the limit of $\omega_D$ as $k \to 0$:

$$\lim_{k \to 0} \omega_D(k) = \frac{M_D}{\Delta \rho}\left[\left(f^* - \frac{J}{M_D}z_0\right)k^2 - \left(\gamma + f^* z_0\right)k^3\right] \qquad 23)$$

At $z_0 = z_{0,c}(J)$, the coefficient of the $k^2$ term in Eq. 23 is zero. This leads to

$$z_{0,c} = \frac{M_D f^*}{J} \qquad 24)$$

which is plotted in Figure 4d. Similar to the interface-controlled case, the shortest wavelength of unstable perturbations is given by $\lambda_{min} = 2\pi\gamma / f^*$ at $z_0 = 0$ regardless of $J$.

We compare the above predictions with experimental observations. In Ref. [12], serrated phase boundaries were found at ~50 nm away from the surface of a chemically delithiated lithium iron phosphate olivine (LiFePO$_4$) sample, which have a periodicity of ~50 nm (see Figure 1c). LiFePO$_4$ is a mainstream cathode material for Li-ion batteries, which undergoes a first-order transition between the LiFePO$_4$ and FePO$_4$ phases upon cycling. Using the orientation-averaged elasticity $E$ (125 GPa [18]), $\nu$ (0.28 [18]), $\epsilon_0$ (2.2% [19]) and $\gamma$ (0.072 J/m$^2$ [20]) from first-principles calculations or experiments, we find the critical wavelength of unstable perturbations at $z_0 = 50$ nm to be 160 nm for interface-controlled or diffusion-controlled intercalation with $J = 0$. In another study [11], oscillatory phase boundaries with a periodicity of ~300 nm is observed at distances of 200 – 400 nm away from the surface of a LiFePO$_4$ single crystal (Figure 1d), at which the critical wavelength is predicted to be between 500 nm and 900 nm. For another common cathode material Li(Ni$_{0.5}$Mn$_{1.5}$)O$_4$, a stripe pattern with a periodicity of ~200 nm was seen in a particle in the Li$_{0.5}$(Ni$_{0.5}$Mn$_{1.5}$)O$_4$ / (Ni$_{0.5}$Mn$_{1.5}$)O$_4$ two-phase region [13], compared to the predicted critical wavelength of 62 nm at $z_0 = 0$ based on the measured or estimated properties of this material [13, 21] ($E = 136$ GPa, $\nu = 0.3$, $\epsilon_0 = 0.06\%$, $\gamma = 0.106$ J/m$^2$). The predictions are in qualitative agreement with the experiments. A more quantitative comparison has to take into account the anisotropic properties of the electrode materials and is also complicated by the fact that the phase boundary keeps moving during the intercalation process.

## 2.2 Phase-field simulation

In this section, we perform numerical simulations using the phase-field method[22, 23] to validate the predictions of the linear stability analysis and study the later stage evolution of the phase boundary morphology not captured by the linear stability theory. In the phase-field model, a scalar field $\phi(\vec{x})$ is used to distinguish between phase I ($\phi = 1$) and phase II ($\phi = 0$), and their interface is represented by the region where $\phi$ varies smoothly from 0 to 1. The free energy of the system is given by

$$F[\phi(\vec{x},t), \epsilon_{ij}(\vec{x},t)] = \int \left[ f_{ch}(\phi) + f_{el}(\epsilon_{ij}, \phi) + \frac{\kappa}{2}(\nabla\phi)^2 \right] dV \qquad 25)$$

where $f_{ch}$ is the homogenous chemical free energy density of the compound and described by a double-well potential

$$f_{ch}(\phi) = \frac{\alpha}{2}\phi^2(1-\phi)^2 + \Delta f_{II \to I} p(\phi) \qquad 26)$$

where $\alpha$ represents the energy barrier between phase I and II, and $p(\phi) = \phi^3(10 - 15\phi + 16\phi^2)$ is a smooth step function that interpolates between $p(0) = 0$ and $p(1) = 1$. For interface-controlled phase transition, $\Delta f_{II \to I}$ represents the volumetric free energy difference between phase I and II in the absence of stress. It is set to 0 for diffusion-controlled phase transition. In simulations, $\alpha$ is chosen to be much larger than $\Delta f_{II \to I}$ and $f^*$ so that the interface energy $\gamma$ and the diffuse interface width $w$ are well approximated by the relations $\gamma = \sqrt{\alpha\kappa}/6$ and $w = 4\sqrt{\kappa/\alpha}$, respectively. $f_{el}(\epsilon_{ij}, \phi) = C_{ijkl}(\epsilon_{ij} - \epsilon_{ij}^0\phi)(\epsilon_{kl} - \epsilon_{kl}^0\phi)/2$ is the linear elastic energy density, where $C_{ijkl}$ is the stiffness tensor and $\epsilon_{ij} = (\partial u_i/\partial x_j + \partial u_j/\partial x_i)/2$ is the elastic strain tensor.

For interface-controlled intercalation, the time evolution of $\phi(\vec{x},t)$ is governed by the Allen-Cahn equation[24]:

$$\frac{\partial\phi}{\partial t} = -\frac{2M_I}{3w}\frac{\delta F}{\delta\phi} \qquad 27)$$

For diffusion-controlled intercalation, it obeys the Cahn-Hilliard equation[25, 26]:

$$\frac{\partial\phi}{\partial t} = \nabla \cdot \left( \frac{M_D}{\rho_0} \nabla \frac{\delta F}{\delta\phi} \right) \qquad 28)$$

in which $\phi$ represents the concentration of $M$ and $\rho_0$ is the molar density of $M$ sites in the compound. The use of a large $\alpha$ ensures that $\phi$ deviates only slightly from 0 or 1 in the bulk of phase I or II so that the assumptions made in the linear stability analysis are respected. Eqs. 27 and 28 reduce to Eqs. 10 and 21 in the sharp interface limit $w \to 0$. Eq. 27 or 28 is solved in conjunction with the linear elasticity equation

$$C_{ijkl}\frac{\partial^2 u_k}{\partial x_j \partial x_l} = C_{ijkl}\epsilon_{kl}^0 \frac{\partial\phi}{\partial x_j} \qquad 29)$$

Simulations are performed for a 2D domain under plane strain condition using the finite-element method implemented in COMSOL Multiphysics 5.3a. The domain size in the $z$ direction is sufficiently large to approximate a semi-infinite system. Traction-free boundary condition is applied to the domain boundary at $z = 0$, and zero displacement and flux boundary conditions are imposed at the opposite boundary. Periodic boundary conditions of $\phi$ and $\vec{u}$ are employed in the

$x$ direction. Parameters used in the simulations are: $\alpha = 1.44 \times 10^{11}$ Pa and $\kappa = 9 \times 10^{-11}$ J m$^{-1}$, which give $\gamma = 0.06$ J m$^{-2}$ and $w = 1$ nm, $E = 100$ GPa, $\nu = 0.3$, $\epsilon_0 = 0.02$, $\rho_0 = \Delta\rho = 25000$ mol m$^{-3}$, $M_I = 2 \times 10^{-16}$ m$^4$ J$^{-1}$ s$^{-1}$ and $M_D = 4.03 \times 10^{-21}$ mol m$^2$ J$^{-1}$ s$^{-1}$.

First, we use simulations to numerically determine $\omega_I(k)$ and $\omega_D(k)$ and compare them to the results of linear stability analysis (Eq. 14 and 22). Specifically, a flat phase boundary with a small sinusoidal perturbation of variable wave vector $k$ is placed at $z_0 = 5$ nm in the initial configuration, which is then let evolve in the absence of intercalation flux ($dz_0/dt = 0$). The time-dependent perturbation amplitude $\delta(t)$ is measured from the phase boundary profile, which is fitted against an exponential function to calculate $\omega_I$ or $\omega_D$. As shown in Figures 3a and 4a, the growth exponents obtained from simulations (red circles) are in good agreement with the linear stability analysis results (solid lines).

Next, we extend the simulations beyond the linear stability regime to investigate the longer term consequence of the stress-induced interface instability. Figure 5a shows the evolution of an originally flat phase boundary perturbed by random noise in the interface-controlled kinetic regime without intercalation flux. As the perturbation amplifies, parts of the phase boundary approach the free surface, which causes phase I to transform from a continuous film into separated domains. Upon the morphological transition, the stress level spikes at the contact points between the boundary and free surface, see Figure 5b, which are likely crack nucleation sites as observed in LiFePO$_4$ [11, 12]. Phase evolution in the diffusion-controlled kinetic regime exhibits similar features, see Figure S1. Figure S2 presents a simulation with a non-zero intercalation flux ($dz_0/dt > 0$). After the moving phase boundary becomes morphologically unstable, the intercalation flux become highly non-uniform, which could lead to detrimental phenomena such as current hotspot and mechanical degradation.

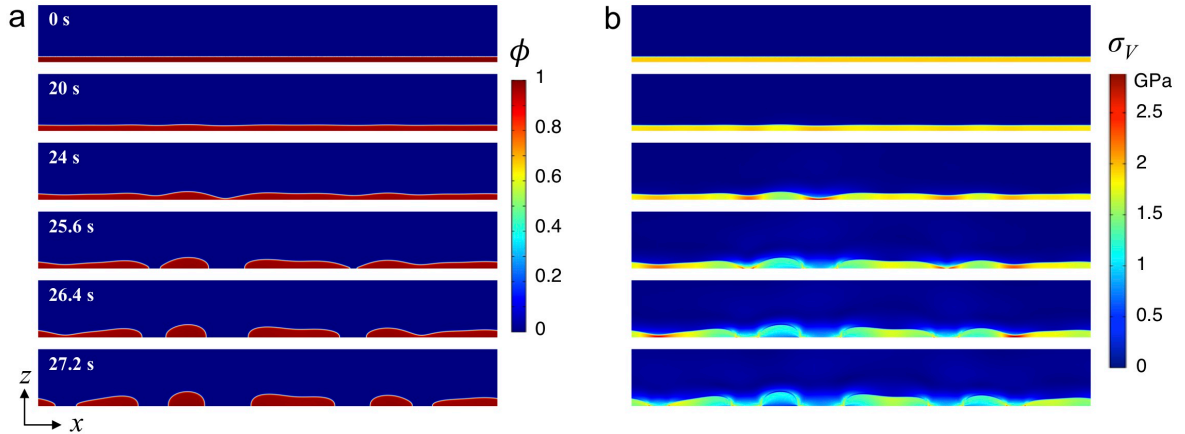

**Figure 5.** Phase-field simulation of the interface-controlled evolution of an initially flat phase boundary located at $z_0 = 5$ nm in the absence of intercalation flux. The boundary is perturbed by a random noise with an amplitude of 0.2 nm at $t = 0$. a) Phase field $\phi$ and b) Von-Mises stress distributions at different times. The computation domain size is 400 nm × 200 nm. The domain is only partially shown along the $z$ axis for clarity.

## 2.3 Stability criterion of uniform intercalation

During intercalation, the interface between the *M*-rich and *M*-poor phases migrates from the surface into the interior of the compound. Simulations presented in the last section show that the interface instability can lead to non-uniform intercalation behavior by causing the product phase to morph into separated individual domains. We ask the question: under what conditions can such morphological transition be suppressed throughout the intercalation process? To derive a tractable stability criterion, here we limit ourselves to linear stability analysis for order-of-magnitude predictions without considering the nonlinear behavior at large $\delta$. Because of its dependence on $z_0$, a moving interface's $\omega_{I/D}$ varies with time and so the perturbation amplitude is given by

$$\delta(k,t) = \delta_0 \exp\left(-\int_0^t \omega_{I/D}(k,z_0(\tau))d\tau\right) \tag{30}$$

For any initially unstable perturbation, its $\omega_{I/D}$ eventually turns negative when $z_0$ becomes sufficiently large. $\delta$ reaches its maximum $\delta_m(k)$ at the interface location $z_0^*(k)$ where $\omega_{I/D}(k,z_0^*) = 0$, i.e. $k$ is the critical wave vector. If the moving phase boundary can maintain its continuity at $z_0^*(k)$ without breaking into individual segments, $\delta$ will decrease afterwards and uniform intercalation will be sustained to the end of the process. Therefore, a stability criterion can be expressed as

$$\delta_m(k) < z_0^*(k) \tag{31}$$

for all *k*.

Now let's consider what Eq. 31 means for a constant-current or galvanostatic battery charging/discharging process, during which the phase boundary moves at a constant velocity *v*: $z_0(t) = z_0(0) + vt$. Let $z_0(0)$ and $\delta_0$ be the thickness of one layer of intercalating ions in the electrode. With change of variable in the integration in Eq. 30, $\delta_m$ can written as $\delta_0 \exp\left[-\int_0^{\omega_{I/D}(k,\delta_0)} (\partial \bar{z}_0 / \partial \omega)_k \omega d\omega / v\right]$, in which $\bar{z}_0(\omega,k)$ is the inverse function of $\omega_{I/D}(\bar{z}_0,k)$ with *k* being held as a constant. Eq. 31 thus becomes

$$v > -\frac{\int_0^{\omega_{I/D}(k,\delta_0)} (\partial \bar{z}_0 / \partial \omega)_k \omega d\omega}{\ln(\bar{z}_0(0,k)/\delta_0)} \equiv v_s(k) \tag{32}$$

On the $v - k$ plane, $v_s(k)$ separates the stable [$v > v_s(k)$] and unstable [$v < v_s(k)$] regions. For interface-controlled intercalation, we use Eq. 14 to evaluate Eq. 32 and find:

$$v_s(k) = M_I \left\{ f^* e^{-2k\delta_0} - \gamma k \left[1 - 2k\delta_0 + \ln\left(\frac{f^*}{\gamma k}\right)\right] \right\} \bigg/ 2\ln\left[\frac{1}{2k\delta_0} \ln\left(\frac{f^*}{\gamma k}\right)\right] \tag{33}$$

For diffusion-controlled intercalation process, $v_s(k)$ needs to be evaluated numerically because of the inter-dependence between $J$ and $v$. Figure 6 shows $v_s(k)$ for both scenarios. An approximate analytical expression of $v_s(k)$ for diffusion-controlled intercalation is derived in Supplementary Note 2 in SI and also shown in Figure 6b as the dashed line, which gives a good estimate to $v_s(k)$. In both cases, it can be seen that uniform intercalation can be achieved if $v$ is above a critical value $v_{s,\max} = \max_k v_s(k)$, which translates to a critical (dis)charging rate.

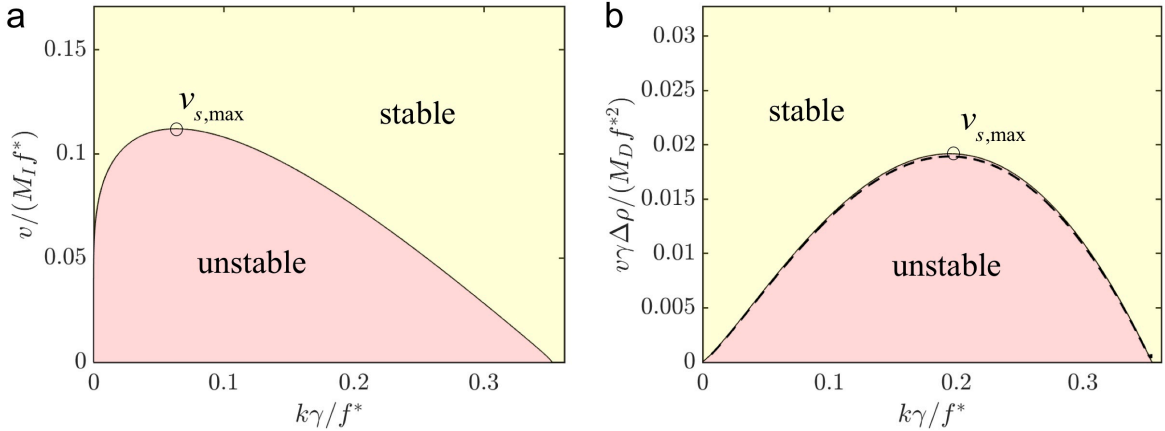

**Figure 6.** Stability of a moving interface at a constant velocity $v$ during a) interface-controlled and b) diffusion-controlled intercalation. $\delta_0 = 1$ nm and $f^*/\gamma = 1.47 \times 10^9$ m$^{-1}$ are assumed. $k$ scaled by $f^*/\gamma$, and $v$ is scaled by $M_I f^*$ in a) and $M_D f^{*2}/\gamma \Delta \rho$ in b). The dashed line in b) is a first-order approximation to $v_s(k)$ (see Supplementary Note 2 in SI).

Determining the numerical value of the critical interface velocity $v_{s,\max}$ requires the knowledge of $M_I$ or $M_D$. There is no report of $M_I$ for any intercalation compounds to our knowledge, but lithium diffusivity data are available for many Li-ion battery electrode materials to allow for an estimate of $v_{s,\max}$ in the diffusion-controlled kinetic regime. Take LiFePO$_4$ as an example, which has Li diffusivity in the range of $10^{-12} - 10^{-11}$ cm$^2$/s [27]. Evaluating its Li mobility from the Einstein relation $M_D = D/RT$, where $R$ is gas constant and $T$ is temperature, using the properties of LiFePO$_4$ (Sec. 2.1) and assuming $\delta_0 = 0.5$ nm, we find $v_{s,\max}$ to be between 8.2 and 82 nm/s at room temperature. For Li(Ni$_{0.5}$Mn$_{1.5}$)O$_4$ with $D_{Li} \approx 10^{-10}$ cm$^2$/s [28], we find $v_{s,\max} \approx 6.5$ nm/s. To put these estimates in perspective, they correspond to fully (dis)charging electrode particles of 1 μm size within 6 s – 80 s, which far exceeds the typical battery (dis)charging rates. Therefore, the onset of stress-induced non-uniform intercalation is likely to be widespread in battery applications.

## 3. Discussion

In this work, we predict the stress-induced intercalation instability and show it to be a practically relevant phenomenon. More studies are needed to elucidate its implications on the performance and degradation of intercalation compounds in different application contexts. Our analysis shows that the instability can be suppressed by fast charging / discharging. Another counterintuitive prediction is that the interface could be made more stable by reducing the diffusivity of intercalation solutes if the phase transition is diffusion-controlled. As the critical interface velocity $v_{s,\max}$ scales linearly with $D$, a 100-fold reduction in the Li diffusivity in LiFePO$_4$ or Li(Ni$_{0.5}$Mn$_{1.5}$)O$_4$ can stabilize the uniform intercalation at typical battery (dis)charging rates. Nevertheless, both high rates and sluggish diffusion will increase the polarization or potential drop within the electrodes and result in inferior capacity utilization due to premature termination of the (dis)charge process. To maintain a stable interface, however, it is not necessary to impose a large intercalation flux through the entire process. As the analysis reveals, a flat phase boundary is most susceptible to instability at the beginning of the (de)intercalation process, when it is near the free surface. An effective mitigation strategy is thus to (dis)charge electrodes at variable rates, first applying a high current pulse to move the interface quickly out of the "unstable zone" and then returning to the lower normal rate to achieve higher capacity. For diffusion-controlled process, the stable zone starts at $z_{0,c}$ given by Eq. 24. For interface-controlled intercalation, even though $k_c$ never reaches zero in a semi-infinite system, practically no perturbations can grow in a finite-size particle when $k_c(z_0) < 2\pi / L$, where $L$ is the lateral particle dimension, i.e. when the particle can no longer accommodate one wavelength of any unstable perturbations.

We discuss several complicating factors not considered in the current analysis. First, coherency stress can be relaxed by the formation of interface dislocations. In an in-situ high resolution transmission electron microscopic (HRTEM) study [29], interface dislocations were found to migrate together with the lithium intercalation front in LiFePO$_4$. This may explain the absence of short-wavelength (< 40 nm) interface oscillations in the sample although longer oscillations cannot be ruled out due to the limited field of view of HRTEM. However, phase boundaries in this study moved at a very low speed (~0.01 nm/s). At typical (dis)charge rates, it is likely that the slow-moving dislocations are unable to keep up with the fast-moving intercalation front and do not contribute to interface stabilization. Second, intercalation can also be surface-reaction-controlled [30, 31] in addition to the interface- and diffusion-controlled mechanisms considered here. The effect of stress on interface stability and its coupling with other intercalation instability mechanisms [32] in this kinetic regime are subject to further study. Third, our analysis only considers systems with isotropic properties. Intercalation compounds often exhibit strong anisotropies in misfit strain, interface energy and/or solute transport, which are expected to further enrich the instability phenomena.

## 4. Conclusion

We performed a linear stability analysis and phase-field simulation of the evolution of a planar coherent phase boundary that is formed during the interface- or diffusion-controlled

(de)intercalation of solute atoms into intercalation compounds, which are widely used as battery electrode or hydrogen storage materials. Results show that the phase boundary is morphologically unstable with respect to the growth of perturbations and non-uniform (de)intercalation behavior ensues from the instability in the presence of a misfit strain between the solute-rich and solute-poor phases. The critical wavelength of unstable perturbations is controlled by the competition between the elastic strain and interface energies, and diverges when the separation between the phase boundary and the free surface of the system increases. For diffusion-controlled intercalation, the instability is also suppressed by increasing intercalation flux and decreasing solute mobility. The predictions are compared with experimental observations of serrated phase boundaries in $LiFePO_4$ and $Li(Ni_{0.5}Mn_{1.5})O_2$ electrode particles with qualitative agreement. While a moving phase boundary during (de)intercalation can maintain its stability when it travels at speeds larger than a critical velocity, this velocity is found to be greater than the charging / discharging rates typically seen in battery applications. Therefore, the stress-induced interface instability in intercalation compounds has practical relevance to the operation of rechargeable batteries.

**Acknowledgments**

YZ and MT are supported by DOE under project number DE-SC0019111. Simulations were partially performed on supercomputers at the Texas Advanced Computing Center (TACC) at The University of Texas and the National Energy Research Scientific Computing Center, a DOE Office of Science User Facility supported by the Office of Science of the U.S. Department of Energy under Contract No. DE-AC02-05CH11231.

Supplementary Information for

**Stress-Induced Intercalation Instability**

Youtian Zhang[1] and Ming Tang[1*]

1. Department of Materials Science & NanoEngineering, Rice University, Houston, TX 77005, USA.

*Corresponding author. Email address: mingtang@rice.edu

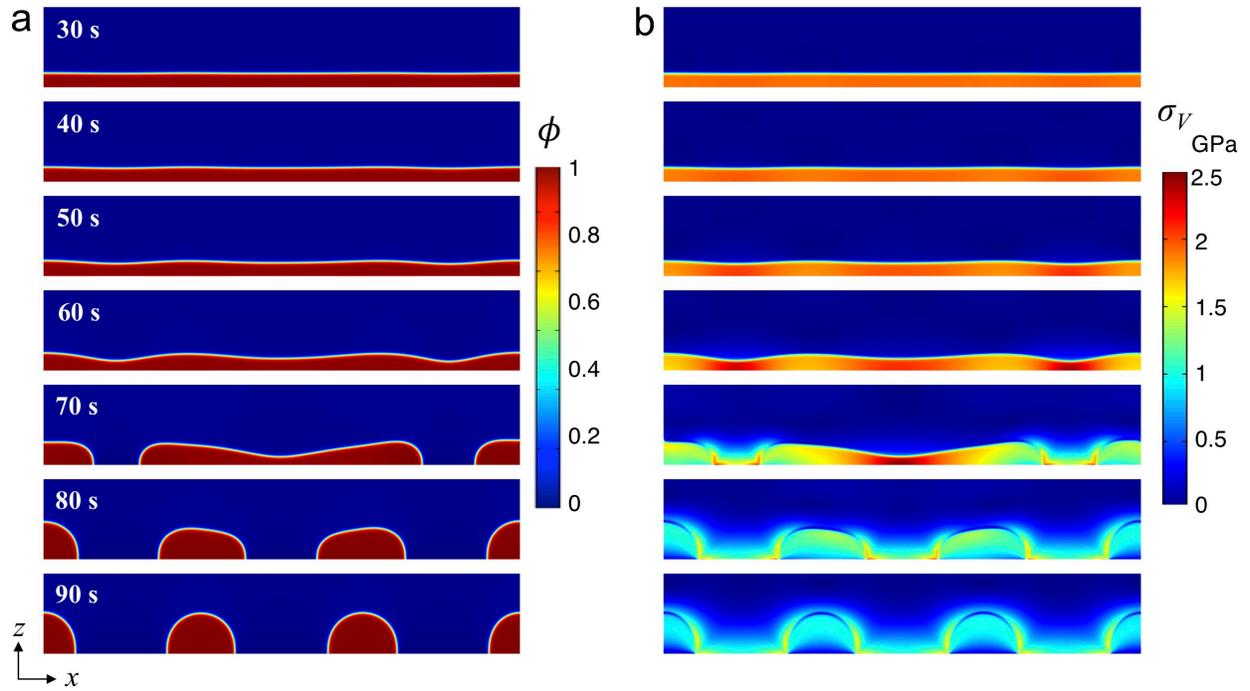

**Figure S1.** Phase-field simulation of the diffusion-controlled evolution of an initially flat phase boundary located at $z_0 = 5$ nm in the absence of an intercalation flux. The boundary is perturbed by a random noise with an amplitude of 0.2 nm at $t = 0$. a) Phase field $\phi$ and b) Von-Mises stress distributions at different times. The computation domain size is 160 nm × 100 nm. The domain is partially shown for clarity. Other simulation parameters are given in the main text.

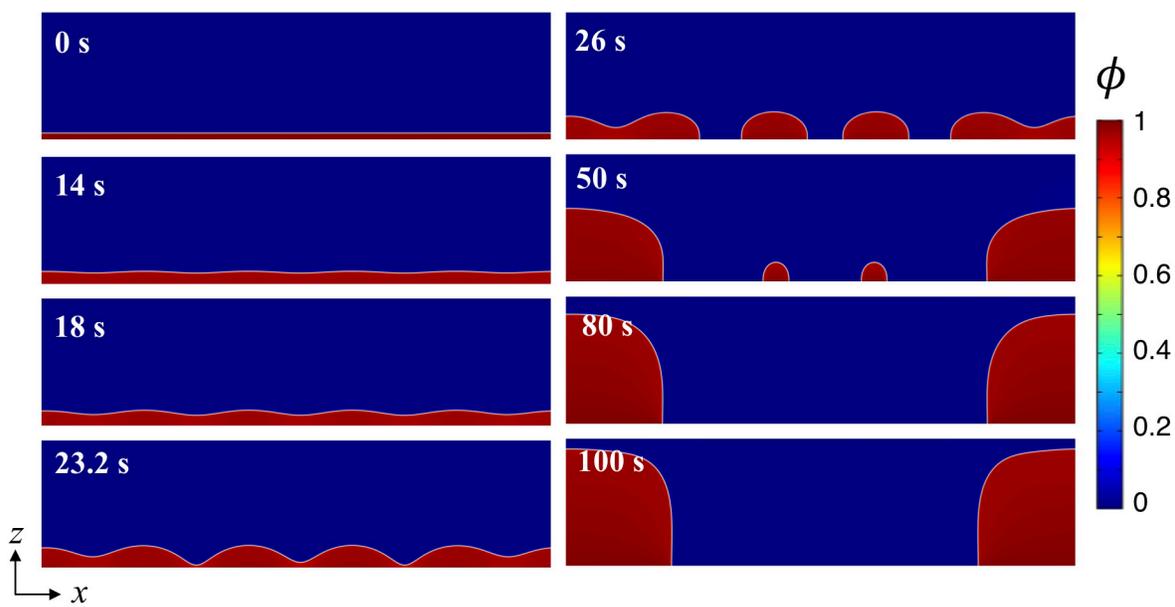

**Figure S2.** Phase-field simulation of phase evolution during interface-controlled intercalation at a constant rate of 5C (system fully intercalated in 720 s). Phase boundary is initially at $z_0 = 5$ nm. Computation domain size is 400 nm × 200 nm and only half of the domain is shown. Other simulation parameters are given in the main text.

**Supplementary Note 1** – Derive the boundary conditions from the displacement continuity at phase boundary

Here we show how the boundary conditions Eqs. 6 and 7, which are expressed in terms of stress components, are derived from the displacement continuity at the interface between phase I and II:

$$u_\tau^I = u_\tau^{II} \qquad \text{S1)}$$

$$u_n^I = u_n^{II} \qquad \text{S2)}$$

$\tau$ and $n$ are the tangential and normal directions of the interface. We define an orthogonal curvilinear coordinate system in the interface region, in which the two axes are always parallel to the interface tangential and normal directions, respectively. As shown in Figure S3, in the neighborhood of a given interface point, this coordinate system overlaps with a circular coordinate system, the origin of which is located at the center of curvature of this point. Let $(r, \theta)$ be the circular coordinate, we have the following relations in this neighborhood

$$u_\tau = u_\theta \qquad \text{S3)}$$

$$u_n = u_r \qquad \text{S4)}$$

$$\epsilon_{\tau\tau} = \epsilon_{\theta\theta} = \frac{1}{r}\frac{\partial u_\theta}{\partial \theta} + \frac{u_r}{r} \qquad \text{S5)}$$

$$\epsilon_{\tau n} = \epsilon_{r\theta} = \frac{1}{2}\left(\frac{1}{r}\frac{\partial u_r}{\partial \theta} + \frac{\partial u_\theta}{\partial r} - \frac{u_\theta}{r}\right) \qquad \text{S6)}$$

Because Eqs. S1 and S2 are always satisfied along the interface, the relations below also hold at the interface:

$$\frac{\partial u_\theta^I}{\partial \theta} = \frac{\partial u_\theta^{II}}{\partial \theta} \qquad \text{S7)}$$

$$\frac{\partial u_r^I}{\partial \theta} = \frac{\partial u_r^{II}}{\partial \theta} \qquad \text{S8)}$$

By inserting Eqs. S2 and S7 into Eq. S5, one finds:

$$\epsilon_{\tau\tau}^I = \epsilon_{\tau\tau}^{II} \qquad \text{S9)}$$

Apply Eqs. S1 and S8 to Eq. S6, and recognize that $\epsilon_{\tau n}^I = \epsilon_{\tau n}^{II}$ at the interface because $\epsilon_{\tau n} = \sigma_{\tau n}/2G$ and shear stress is continuous across the interface, which lead to:

$$\frac{\partial u_\theta^I}{\partial r} = \frac{\partial u_\theta^{II}}{\partial r} \qquad \text{S10)}$$

Since Eq. S10 holds for any point on the interface, the above identity still holds after differentiation with respect to $\theta$:

$$\frac{\partial}{\partial r}\left(\frac{\partial u_\theta^I}{\partial \theta}\right) = \frac{\partial}{\partial r}\left(\frac{\partial u_\theta^{II}}{\partial \theta}\right) \qquad \text{S11)}$$

Substituting Eq. S5 into the above equation, one has:

$$\frac{\partial}{\partial r}\left(r\epsilon_{\theta\theta}^{I} - u_{r}^{I}\right) = \frac{\partial}{\partial r}\left(r\epsilon_{\theta\theta}^{II} - u_{r}^{II}\right) \quad \text{S12)}$$

Because of Eq. S9 and $\epsilon_{rr} = \partial u_{r}/\partial r$, Eq. S12 can be rewritten as:

$$\frac{\partial \epsilon_{\theta\theta}^{I}}{\partial r} - \frac{\epsilon_{rr}^{I}}{r} = \frac{\partial \epsilon_{\theta\theta}^{II}}{\partial r} - \frac{\epsilon_{rr}^{II}}{r} \quad \text{S13)}$$

or

$$\frac{\partial \epsilon_{\tau\tau}^{I}}{\partial n} - \kappa\epsilon_{nn}^{I} = \frac{\partial \epsilon_{\tau\tau}^{II}}{\partial n} - \kappa\epsilon_{nn}^{II} \quad \text{S14)}$$

where $\kappa = 1/r$ is the radius of curvature of the local interface segment. Therefore, the displacement continuity condition (Eqs. S1 and S2) are replaced by two boundary conditions in terms of the elastic strain tensor (Eqs. S9 and S14). To express the boundary conditions in terms of the stress tensor, we employ the Hooke's law under plane strain condition:

$$\epsilon_{\tau\tau}^{I} - \epsilon_{0} = \frac{1+v}{E}\left[(1-v)\sigma_{\tau\tau}^{I} - v\sigma_{nn}^{I}\right] \quad \text{S15a)}$$

$$\epsilon_{\tau\tau}^{II} = \frac{1+v}{E}\left[(1-v)\sigma_{\tau\tau}^{II} - v\sigma_{nn}^{II}\right] \quad \text{S15b)}$$

$$\epsilon_{nn}^{I} = \frac{1+v}{E}\left[(1-v)\sigma_{nn}^{I} - v\sigma_{\tau\tau}^{I}\right] \quad \text{S15c)}$$

$$\epsilon_{nn}^{II} = \frac{1+v}{E}\left[(1-v)\sigma_{nn}^{II} - v\sigma_{\tau\tau}^{II}\right] \quad \text{S15d)}$$

By applying Eq. S15 to Eqs. S9 and S14, we obtain Eqs. 6 and 7 in the main text.

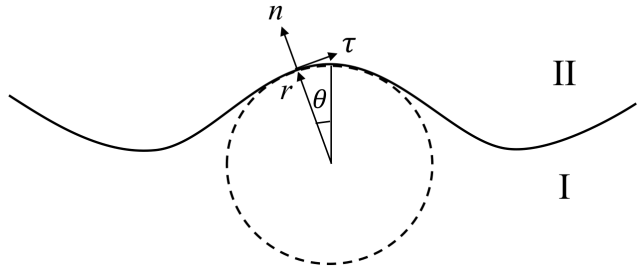

**Figure S3.** The orthogonal curvilinear coordinate system in the neighborhood of an interface point is equivalent to a circular coordinate system whose origin is located at the center of curvature of this point.

**Supplementary Note 2** – Stability criterion of uniform intercalation for diffusion-controlled process

For diffusion-controlled intercalation, the inverse function of $\omega_D(\bar{z}_0, k)$ with $k$ being held constant can derived from the dispersion relation in Eq. 22:

$$\bar{z}_0(\omega, k) = \frac{1}{k} \operatorname{arctanh}\left( \frac{M_D k(f^* - \gamma k) - \omega \Delta \rho / k}{M_D k(f^* + \gamma k) + J} \right) \quad \text{S16)}$$

Substituting Eq. S16 into Eq. 32 in the main text and replacing $J$ with $\Delta \rho v$, we obtain an equation satisfied by the critical velocity $v_s(k)$:

$$v_s(k) = \left\{ (2M_D f^* k + \Delta \rho v_s(k)) \ln\left[ \frac{2M_D f^* k + \Delta \rho v_s(k)}{(M_D k(f^* + \gamma k) + \Delta \rho v_s(k))(1 + \tanh(k\delta_0))} \right] \right.$$
$$\left. + (2M_D \gamma k^2 + \Delta \rho v_s(k)) \ln\left[ \frac{2M_D \gamma k^2 + \Delta \rho v_s(k)}{(M_D k(f^* + \gamma k) + \Delta \rho v_s(k))(1 - \tanh(k\delta_0))} \right] \right\} \quad \text{S17)}$$
$$\bigg/ 2\Delta\rho \ln\left[ \frac{1}{k\delta_0} \operatorname{arctanh}\left( \frac{M_D k(f^* - \gamma k)}{M_D k(f^* + \gamma k) + \Delta \rho v_s(k)} \right) \right]$$

Eq. S17 gives an implicit solution to $v_s(k)$, whose value needs to be determined numerically. A zeroth-order analytical approximation, $v_s^0(k)$, is derived by letting $v_s(k) = 0$ on the right hand side of Eq. S17:

$$v_s^0(k) = M_D f^* k \frac{\ln\left[ \frac{2f^*}{(f^* + \gamma k)(1 + \tanh(k\delta_0))} \right] + \frac{\gamma k}{f^*} \ln\left[ \frac{2\gamma k}{(f^* + \gamma k)(1 - \tanh(k\delta_0))} \right]}{\Delta \rho \ln\left[ \frac{1}{k\delta_0} \operatorname{arctanh}\left( \frac{f^* - \gamma k}{f^* + \gamma k} \right) \right]} \quad \text{S18)}$$

Figure S4 compares $v_s^0(k)$ with the exact $v_s(k)$. A better first-order approximation, $v_s^1(k)$, is obtained by replacing $v_s(k)$ on the right hand side of Eq. S17 with $v_s^0(k)$. Figure S4 shows that $v_s^1(k)$ provides a good approximation to the exact $v_s(k)$.

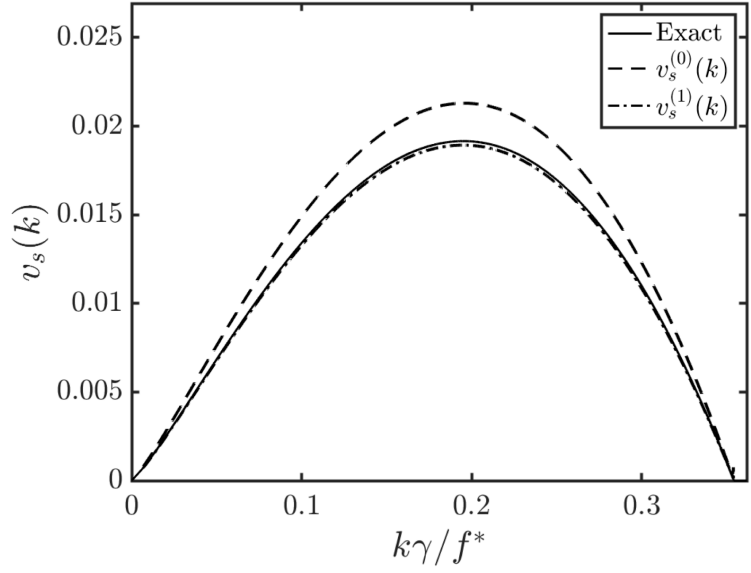

**Figure S4.** Comparison between $v_s(k)$ and its zeroth-order [$v_s^0(k)$] and first-order [$v_s^1(k)$] approximations for diffusion-controlled intercalation.